# Two cellular resource-based models linking growth and parts characteristics aids the study and optimisation of synthetic gene circuits

*Huijuan Wang[1], Maurice H.T. Ling[1], Tze Kwang Chua[1], Chueh Loo Poh[2]* ✉

[1]*School of Chemical and Biomedical Engineering, Nanyang Technological University, Singapore 637459, Singapore*
[2]*Department of Biomedical Engineering, National University of Singapore, Singapore 117583, Singapore*
✉ *E-mail: poh.chuehloo@nus.edu.sg*

**Abstract:** A major challenge in synthetic genetic circuit development is the inter-dependency between heterologous gene expressions by circuits and host's growth rate. Increasing heterologous gene expression increases burden to the host, resulting in host growth reduction; which reduces overall heterologous protein abundance. Hence, it is difficult to design predictable genetic circuits. Here, we develop two biophysical models; one for promoter, another for RBS; to correlate heterologous gene expression and growth reduction. We model cellular resource allocation in *E. coli* to describe the burden, as growth reduction, caused by genetic circuits. To facilitate their uses in genetic circuit design, inputs to the model are common characteristics of biological parts [e.g. relative promoter strength (RPU) and relative ribosome binding sites strength (RRU)]. The models suggest that *E. coli*'s growth rate reduces linearly with increasing RPU/RRU of the genetic circuits; thus, providing 2 handy models taking parts characteristics as input to estimate growth rate reduction for fine tuning genetic circuit design *in silico* prior to construction. Our promoter model correlates well with experiments using various genetic circuits, both single and double expression cassettes, up to a relative promoter unit of 3.7 with a 60% growth rate reduction (average $R^2 \sim 0.9$).

## 1 Introduction

Synthetic biology has been defined as the engineering of biology involving the synthesis of complex, biologically based (or inspired) systems which display functions that do not exist in nature [1]. A number of studies have been reported for various applications having special potentials or promising breakthroughs in human health [2, 3], environment [4] or biotechnology [5, 6], all involving foreign genetic circuits to control cellular activities [7]. A successfully designed genetic circuit not only relies on the engineering of parts such as promoters and gene sequence but also the ability to link parts together to achieve a predictable behaviour [8]. However, it remains a challenge in synthetic biology to design and build genetic circuits in a predictable manner [9]. This is largely due to the complexity of biological systems and inter-dependency between genetic circuits and their hosts [8, 10–13]. Such inter-dependencies can be represented by reduced growth rate upon heterologous gene expression, resulting in a growth-rate dependent abundance of total heterologous proteins [10, 11]. It has been demonstrated that growth rate was inversely related to exogenous plasmid size and copy number, as well as the strength of promoters within the genetic circuits [14–18]. Since growth rate determines cell number, reduced growth rate affects total heterologous protein abundance, hence, affecting the quantity of harvestable heterologous protein, under the assumption that the proportion of heterologous protein per cell is constant.

This inter-dependency could be due to the sharing of cellular resources between synthetic genetic circuits and the host microbes. These resources include construction materials (e.g. amino acids, nucleotides etc.) and energy (e.g. ATP) in the host microbes [19]. Here, we denote energy as the effort required to put the construction materials together to form the cellular components (e.g. protein, RNAs etc.). Since the host's natural processes (metabolism, cell division etc.) require cellular resources, the uptake of these cellular resources by 'foreign' genetic circuits is likely to result in resource stress, leading to reduction in growth rate [15].

To better understand inter-dependency caused by cellular resource allocation, several mathematical models have been proposed. Bentley *et al.* [16] elaborated metabolic burden, in terms of growth rate variation, experienced by engineered bacterium. In this model, growth rate was affected by the variations of eight types of cellular components. However, this model is difficult to use as more than ten parameters are required. To balance the uncertainty caused by 'too many' parameters, Carrera *et al.* [20] utilised an established host-cell model which offered parameter values of some cellular components such as DNA polymerase, RNA polymerase and ribosomes. As these cellular components are used to express heterologous proteins, Carrera *et al.* [20] derived and validated a phenomenological model describing the dependency of these cellular components on growth rate in engineered bacterium, and validated the model with experimental data using different media and plasmid copy numbers. However, its major drawback is that the model is insufficiently validated with the genetic circuit designs beyond one constitutive transgenic expression. In addition, Carrera *et al.* [20] suggest that ribosomes are the limiting factor among the considered resource, which is corroborated by Scott *et al.* [11]. After discussing the proteome fraction allocated to host protein and heterologous protein synthesis, Scott *et al.* [11] proposed a linear relationship between growth rate and protein production. As only few parameters are involved, this work is acknowledged as a quantitative prediction of a priori growth rate. However, Scott *et al.*'s [11] model was mainly validated using Isopropyl β-D-1-thiogalactopyranoside (IPTG) induction. Since IPTG induction results in bimodal gene expression in a few conditions, it possibly obscures the results by comprising cells induced by IPTG in different ways [21]. To remove the potentially clouding effect caused by IPTG-induced validation, Bienick *et al.* [17] considered unimodal gene expression, in terms of constitutive heterologous expression, to test the growth rate variation along with the allocation model of ribosome [17]. Combined, ribosome-based resource allocation models have been proposed to study the growth reduction caused by single heterologous expression. However, the usage of transcriptional resource, as well as the model adaptability in complex synthetic circuits, has not been discussed in previous studies. Despite the common theme of ribosome allocation in various models, the estimation of ribosome allocation to genetic circuits is not easily deduced experimentally and can be confounded by ribosome recycling [22, 23]. Yet, a number of studies suggest that cellular resources are optimally allocated to maximise growth rate in a given condition [24, 25], even between growth requirements and the maintenance of cellular machineries [26], implying that resource allocations may be abstracted by simple partitioning heuristics [27].

To investigate whether a model based on resource allocation partitioning heuristics could predict the impact on the growth of the host due to synthetic genetic circuits, we developed two biophysical models in this study, one for the promoter and one for the ribosome binding site (RBS). The models describe the partitioning of cellular resources (used in protein, DNA, RNA and lipid syntheses) between host *E. coli* cells and genetic circuits, under the assumption of optimal resource allocation between endogenous requirements and genetic circuits. Using relative RBS strength, our RBS model predicts a linear growth reduction on host *E. coli* cells resulted from an increased RBS strength in heterologous expression under the same promoter. Using relative promoter strength, our promoter model predicts a linear growth reduction on host *E. coli* cells resulted from increased promoter strength in both single and double heterologous expressions under the same RBS, demonstrating an additive effect of promoter strengths on growth rate. Hence, this study provides two handy models, which take standard parts characteristics [28–30] as input, to estimate growth rate reduction for the design of synthetic genetic circuits.

## 2 Model formation

In this study, we first developed a biophysical model of the host, by reference to a single host cell, based on the balance between the nutrient's (i.e. glucose) uptake rate ($V_{\text{uptake}}^H$, molecules of glucose per hour) and the nutrient's consumption rate of the host system ($V_{\text{consumption}}^H$, molecules of glucose per hour) and exogenous plasmids ($V_{\text{consumption}}^P$, molecules of glucose per hour) during exponential growth [31] (1). In this study, we considered *E. coli* as the host

$$V_{\text{uptake}}^H = V_{\text{consumption}}^H + V_{\text{consumption}}^P \quad (1)$$

On the assumption that the glucose transporter is evenly distributed on the cell membrane, the glucose uptake rate ($V_{\text{uptake}}^H$) is proportional to the surface area of the cell. *E. coli* is known to be a rod-shaped bacterium with relatively constant density and diameter [32]; hence, its surface area is proportional to its mass (Fig. S1). Phillips *et al.* [19] proposed that *E. coli*'s glucose uptake rate is consistent at $2 \times 10^{21}$ glucose molecules per gram of cells per hour in M9 medium. Hence,

$$V_{\text{uptake}}^H = 2 \times 10^{21} \times M^H \quad (2)$$

where $M^H$ is the host *E. coli* mass (in grams).

Glucose is used by the host to synthesise biomass ($M$), as well as the energy required for synthesising the biomass ($E$, molecules per hour). Phillips *et al.* [19] considered four major classes of cellular components (i.e. DNA, RNA, protein and lipid), which account for more than 98% of the entire biomass of the cell (Table S1). Hence, glucose consumption is denoted as the summation of biomass synthesis and its corresponding energy cost, across the four major classes, as

$$V_{\text{consumption}}^H = \sum_i^{\{\text{protein,lipid,RNA,DNA}\}} (M_i + E_i) \quad (3)$$

At the same time, Phillips *et al.* [19] show that there is a fixed proportion ($\varphi_i$, Table S1) between energy and material cost (4)

$$M_i = \varphi_i \times E_i \quad (4)$$

Hence, by substituting (4) into (3), we reformatted (3) into (5), which focuses on biosynthesis energy cost $E_i$

$$V_{\text{consumption}}^H = \sum_i^{\{\text{protein,lipid,RNA,DNA}\}} (\varphi_i + 1) E_i \quad (5)$$

The $E_i$ (molecules per hour) has been modelled by Wagner [33] as

$$E_i = \varepsilon \times c_i \times a_i \quad (6)$$

where $\varepsilon$ converts the number of glucose molecules needed to produce a molecule of ATP, which is set at 0.03; $c_i$ refers to the energy cost constant (dimensionless, Table S1); and $\alpha_i$ refers to the biosynthesis rate of the corresponding cellular component (molecules of cellular component per hour, Table S1). This gives rise to four cellular components' synthesis rates to consider – DNA ($\alpha_{\text{DNA}}$), RNA ($\alpha_{\text{RNA}}$), protein ($\alpha_{\text{protein}}$) and lipid ($\alpha_{\text{lipid}}$).

We assume that DNA synthesis is negligible ($\alpha_{\text{DNA}} = 0$) compared with other components, and mRNA generally have a short half-life (Bionumber 106869) compared with rRNA and tRNA [34]. As a result, we grouped rRNA and tRNA together as RNA. Hence, only RNA ($\alpha_{\text{RNA}}$, number of RNA molecules synthesised per hour), protein ($\alpha_{\text{protein}}$, number of protein molecules synthesised per hour) and lipid ($\alpha_{\text{lipid}}$, number of lipid molecules synthesised per hour) are considered. To deduce an expression for biosynthesis rate ($\alpha_i$) in terms of quantity of the respective cellular component, we used an ordinary differential equation model (7) and assume that it reaches steady state ($dP_i/dt = 0$) during exponential growth

$$\frac{dP_i}{dt} = \alpha_i - (\beta_{\text{deg}} + \beta_{\text{dil}}) P_i \quad (7)$$

where $P_i$ is the quantity of the cellular component $i$ (number of molecules), $\beta_{\text{deg}}$ refers to degradation rate (per hour) as determined by the cellular component's half-life (Table S1) and $\beta_{\text{dil}}$ refers to dilution rate (per hour) which is equal to growth rate ($\mu$, per hour). For example, the synthesis of protein ($P_{\text{protein}}$) can be modelled as (7a), using (7) where $i = \text{protein}$

$$\frac{dP_{\text{protein}}}{dt} = a_{\text{protein}} - (\beta_{\text{deg}} + \beta_{\text{dil}}) P_{\text{protein}} \quad (7a)$$

As the half-life of most cellular components is much longer than doubling time (Table S1), we simplify (7) into

$$\frac{dP_i}{dt} = a_i - \beta_{\text{dil}} P_i \quad (8)$$

At steady state, (8) can be simplified into (9) as dilution rate ($\beta_{\text{dil}}$, per hour) corresponds to growth rate ($\mu$, per hour)

$$a_i = \beta_{\text{dil}} P_i = \mu P_i \quad (9)$$

Hence, the synthesis rates of RNA ($\alpha_{\text{RNA}}$), protein ($\alpha_{\text{protein}}$) and lipid ($\alpha_{\text{lipid}}$) can be modelled as

$$a_{\text{protein}} = \beta_{\text{dil}} P_{\text{protein}} = \mu P_{\text{protein}} \quad (9a)$$

$$a_{\text{RNA}} = \beta_{\text{dil}} P_{\text{RNA}} = \mu P_{\text{RNA}} \quad (9b)$$

$$a_{\text{lipid}} = \beta_{\text{dil}} P_{\text{lipid}} = \mu P_{\text{lipid}} \quad (9c)$$

The quantity of each cellular component, $P_i$, is modelled separately. As Phillips et al. [19] suggest a fixed proportion of protein ($\sigma_{\text{protein}}$) and RNA ($\sigma_{\text{RNA}}$) in E. coli, the quantity (number of molecules) of protein ($P_{\text{protein}}$) and RNA ($P_{\text{RNA}}$) can be modelled as

$$P_{\text{protein}} = \frac{\sigma_{\text{protein}} \times M^H}{\text{MW}_{\text{protein}}} \times \text{NA} \quad (10)$$

$$P_{\text{RNA}} = \frac{\sigma_{\text{RNA}} \times M^H}{\text{MW}_{\text{RNA}}} \times \text{NA} \quad (11)$$

where $\text{MW}_{\text{protein}}$ and $\text{MW}_{\text{RNA}}$ are the molecular weights of protein and RNA (kilodaltons, Table S1), respectively, and NA refers to Avogadro's constant, which converts the concentration of cellular component (molar) into quantity (number of molecules).

Since a lipid is abundant on cellular membrane, the amount of lipid is proportional to bacterial surface area. Each of the inner and outer membranes consist of a lipid bilayer; hence, the amount of possible lipid coverage can be estimated as four times of the bacterial surface area. However, Phillips et al. [19] suggest that only about 50% of the surface area is covered by lipids; we can estimate the number of lipid molecules ($P_{\text{lipid}}$) based on the surface area of the cell ($A^H$, $\mu m^2$ of cell surface) and the average area of a lipid molecule ($A_{\text{lipid}}$, $\mu m^2$ of cell surface, Table S1)

$$P_{\text{lipid}} = \frac{4 \times 0.5 \times A^H}{A_{\text{lipid}}} \quad (12)$$

Similarly, for exogenous plasmids in engineered E. coli, we assume negligible resources for plasmid replication and mRNA synthesis due to its short half-life [34]. Hence, the glucose consumption rate by exogenous plasmids ($V^P_{\text{consumption}}$, molecules of glucose per hour) is mainly contributed by transgenic protein expression, and can be modelled as

$$V^P_{\text{consumption}} = M^P_{\text{protein}} + E^P_{\text{protein}} \quad (13)$$

Using (4), (13) can be rewritten as

$$V^P_{\text{consumption}} = (\varphi^P_{\text{protein}} + 1) E^P_{\text{protein}} \quad (14)$$

where the $E^P_{\text{protein}}$ (number of glucose molecules/hour) is modelled as (15), using (6) [33]

$$E^P_{\text{protein}} = \varepsilon \times c^P_{\text{protein}} \times \alpha^P_{\text{protein}} \quad (15)$$

The biosynthesis rate of transgenic protein ($\alpha^P_{\text{protein}}$) can be rewritten as (16), using the same principles described in (7)–(9), where $\mu_{\text{promoter}}$ is the growth rate (per hour) of the engineered cell

$$\alpha^P_{\text{protein}} = \mu_{\text{promoter}} \times P^P_{\text{protein}} \quad (16)$$

As the expression of transgenic protein is dependent on the type of promoter, the amount of expressed transgenic protein or heterologous protein is proportional to the strength of the promoter. Hence, (16) can be rewritten as

$$\alpha^P_{\text{protein}} = \mu_{\text{promoter}} \times P^P_{\text{promoter}} \quad (17)$$

Using a standard promoter, Kelly et al. [29] (Fig. S2) defined relative promoter unit (RPU, dimensionless) as

$$\text{RPU} = \frac{\mu_{\text{promoter}} \times P^P_{\text{promoter}}}{\mu_{\text{standard\_promoter}} \times P^P_{\text{standard\_promoter}}} \quad (18)$$

where $\mu_{\text{promoter}}$ and $\mu_{\text{standard\_promoter}}$ are the growth rates (per hour) of the host when $P^P_{\text{promoter}}$ and $P^P_{\text{standard\_promoter}}$ are used for transgenic protein or reporter protein expression, respectively. Hence, RPU is a ratio of growth and protein expression between the promoter of interest ($P^P_{\text{standard\_promoter}}$) and a standard promoter ($P^P_{\text{standard\_promoter}}$). Re-arranging (18) gives

$$P^P_{\text{promoter}} = \frac{\text{RPU} \left( \mu_{\text{standard\_promoter}} \times P^P_{\text{standard\_promoter}} \right)}{\mu_{\text{promoter}}} \quad (19)$$

By multiplying growth rate ($\mu_{\text{promoter}}$), (19) can be linked to (17)

$$\alpha^P_{\text{protein}} = \mu_{\text{promoter}} \times P^P_{\text{promoter}}$$
$$= \text{RPU} \left( \mu_{\text{standard\_promoter}} \times P^P_{\text{standard\_promoter}} \right) \quad (20)$$

Besides a promoter, a RBS has also been used to modulate transgenic expression [30, 35]. Hence, it is plausible to consider the RBS's counterpart to RPU – a dimensionless relative RBS unit (RRU) – and by substituting promoter and RPU with RBS and RRU, respectively, into (17)–(19), an RRU/RBS counterpart of (20) can be formulated as

$$\alpha^P_{\text{protein}} = \mu_{\text{RBS}} \times P^P_{\text{RBS}} = \text{RRU} \left( \mu_{\text{standard\_RBS}} \times P^P_{\text{standard\_RBS}} \right) \quad (21)$$

By combining (2)–(21) into (1) (described in Description S1), the growth reduction as a result of heterologous protein expression can be estimated by

$$\mu_{\text{engineered}} = \mu_{\text{WT}} - 0.16 \times \text{RPU} \quad (22a)$$

$$\mu_{\text{engineered}} = \mu_{\text{WT}} - 0.16 \times \text{RRU} \quad (22b)$$

where $\mu_{\text{WT}}$ and $\mu_{\text{engineered}}$ are the growth rates (per hour) of the wild type and engineered E. coli, respectively.

Equations (22a) and (22b) suggest that linearly reduced growth rate ($\mu$) is caused by an increase of RPU or RRU, respectively. In this context, RPU and RRU represent high-level, coarse-gained partitioning of transcriptional and translational resources for heterologous expression under the assumption of optimal resource allocation [24, 25]. However, genetic circuits can express two or more heterologous proteins transcribed by the same or different promoters. To model more complex genetic circuits, we assume that those heterologous expressions are independent. Hence, glucose consumption caused by each heterologous expression is orthogonal and additive. Thus, (23) can be reformulated as

$$\mu_{\text{engineered}} = \mu_{\text{WT}} - 0.16 \times \sum_{i}^{N} \text{RPU}_i \quad (23)$$

## 3 Materials and method

### 3.1 Strains and plasmids

E. coli MG1655 (ATCC 700926) was used as a host for cloning and expression of our constructed plasmids. Pseudomonas aeruginosa PAO 1 was used as the genome DNA template for the cloning of lasR-plasI quorum sensing (QS) system. Plasmids conforming to BglBrick standard [36]: pBbE2k, pBbE6K and pBbE8K were supplied by Addgene (Cambridge, USA).

### 3.2 System assembly

Constitutive promoters (Table S2), *plasI* and *lasR* gene were cloned from an earlier study [37]. The sequence of ribosome binding site (community RBS collection) and terminator (BBa_B0015) were taken from standard parts registry [38]. We chose *rrnB P1* promoter [39], RBS from pBb backbone (sequence: TTTAAGAAGGAGA TATACAT), and terminator (BBa_B0015) as the standard constitutive promoter, RBS and terminator, respectively. Red fluorescence protein (RFP) was used as the reporter protein. For system assembly, all PCR was performed using iProof PCR kit (BioRad, USA) in a dual cycle PCR program. Biological parts were spliced using New England Biolabs double digestion enzymes (NEB, Canada) and ligated to vectors pBbE8K using T4 DNA ligase (NEB, Canada). Composite systems with two or more biological modules were sequentially assembled.

### 3.3 Media and growth

M9 medium (glucose +) was prepared with M9 salt (Sigma-Aldrich, USA), supplemented with 0.4% (v/v) casamino acids (Sigma-Aldrich, USA), 0.2 mg/l thiamine hydrochloride (Sigma-Aldrich, USA) and 20 mM D-glucose (Sigma-Aldrich, USA). For glucose update study, M9 medium (2-NBDG +) was prepared following a similar protocol but replacing glucose by 20 mM 2-(N-(7-nitrobenz-2-oxa-1,3-diazol-4-yl)-amino)-2-deoxyglucose (2-NBDG, life technologies, USA). 2-NBDG is a glucose analogue [40] and a fluorescence dye with excitation 485 nm and emission 525 nm. It has been shown that uptake rate of 2-NBDG by *E. coli* equals to uptake rate of D-glucose [41, 42].

### 3.4 Glucose uptake rate measurement

Glucose uptake rate during exponential growth phase was measured on wild type and engineered *E. coli*, which carried pBbE8K plasmids with 14 constitutive promoters, respectively. In brief, cells grown to stationary phase in Luria-Bertani (LB) medium were diluted 100× in fresh LB medium. The diluted cells were grown to early exponential growth phase ($OD_{600} = 0.15$–0.3) [43], before replacing the growth medium with M9-2NBDG medium and subjected to microplate fluorescence assay. The cell cultures were measured for their 2-NBDG fluorescence for 4 h at a fixed interval of 10 min. Fluorescence activity of 2-NBDG was measured using Synergy HT microplate reader (Bio-Tek, USA) at 485 nm excitation and 525 nm emission. Wild type *E. coli* MG1655 was used as negative control. Three biological replicates were used.

### 3.5 Synthetic biological parts characterisation

RPU were characterised on engineered *E. coli*, which carried pBbE8K plasmids ligated with genetic circuits comprising constitutive promoter, standard RBS and RFP. Cells were first grown to stationary phase in LB medium and then diluted 100× in fresh LB medium. The diluted cells were grown to early exponential growth phase ($OD_{600} = 0.15$–0.3), before replacing the growth medium to M9 (glu+) and then subjected to microplate fluorescence assay. Fluorescence activities of RFP were measured using Synergy HT microplate reader (Bio-Tek, USA) at 535 nm excitation and 600 nm emission. Meanwhile, absorbance measurements were taken at 600 nm, as an indicator for the population density of *E. coli* in 300μl sample volume. Wild type *E. coli* MG1655 was used as negative control. Three biological replicates were used. RRUs were characterised on engineered *E. coli*, which carried pBbE8K plasmids ligated with genetic circuits comprising a standard constitutive promoter (rrnB P1), different RBS and RFP. Same measurement protocol other than promoter study was applied.

### 3.6 QS circuits assembly and characterisation

LasR-*plasI* system was designed and published in our previous work [37]. In this study, we replaced the constitutive promoter *csiDp* with constitutive 70σ promoters: BBa_J23105, BBa_J23108 and *rrnB P1* which represent weak, medium and strong promoters, respectively. Similar characterisation protocol was used in the three LasR-*plasI* systems with extra inducer: 3OC12HSL (AHL, Sigma-Aldrich, USA) was added and cell cultures were subjected to microplate assay. Final concentration of 3OC12HSL varied from $10^{-10}$ to $10^{-5}$ M and RFP fluorescence plus absorbance at 600 nm were measured for 6 h at a fixed interval of 10 min. Wild type *E. coli* MG1655 was used as negative control. Three biological replicates were used.

### 3.7 RFP quantification

RFP fluorescence intensity correlates to protein abundance. To quantify RFP concentration, engineered *E. coli* which constitutively expressed RFP driven by the reference promoter (*rrnB P1*) was used. Cell culture grown to stationary phase was pelleted at 4000 g for 10 min and washed twice using 1× phosphate buffered saline. The re-suspend cell pellet was sonicated for 10 min with a 10 s oscillation (Branson Digital Sonifier, USA). After filtering the cell mixture using a 0.2 μM Supor membrane filter (Pall Life Science, USA), 0.5 ml of the filtrate was loaded in fast protein liquid chromatography (FPLC) (AKTA Pure, GE healthcare, USA) using HI prep 16/60 Sephacryl S 100 HR column with a flow rate of 0.5 ml/min. RFP sample tubes were then chosen based on the chromatogram from FPLC assay (Fig. S1). To quantify RFP concentration, a Quant-IT protein assay kit (Life Technologies, USA) was used, with the assay procedure provided by the manufacturer.

### 3.8 Data analysis and statistics

All data were presented by mean ± SD. Model simulation and sensitivity analysis were conducted by Matlab R2012a (Mathwork, USA). Correlation between simulated and experimental values is represented by Pearson's correlation. To calculate the glucose uptake rate, area under curve was approximated using Simpson's rule with 128 intervals. Chi-square test was applied to study the significance between glucose uptake rates of engineered *E. coli*.

## 4 Results and discussion

The developed model was validated using experimental results. Our results show good correlation between the predicted and observed growth rates and there was total heterologous protein abundance from various biological parts and circuit designs.

### 4.1 Strong correlation between predicted and observed growth reduction in a single expression cassette

We constructed a set of genetic circuits expressing heterologous RFP driven by various promoters or RBSs, and measured the impact of these parts on the host in terms of growth reduction. Nine RBSs with RRUs ranging from 0.01 to 1 (Fig. 1*a*) and 14 constitutive promoters with RPUs ranging from 0.01 to 1.83 (Fig. 1*b*, Table S2) were used. The parts were characterised using a reference characterisation genetic circuit: vector backbone-pBbE8K, promoter – *rrnB P1*, RBS from pBbE8K, reporter protein – RFP under the following characterisation media: M9 medium with 20 mM glucose. Host growth rate during the exponential growth phase was measured and compared with the growth rate suggested by our model. Our results show that experimental results from engineered *E. coli* agree with model prediction (Fig. 1*c*, $R^2 = 0.86$, $n = 23$, *p*-value < 0.01) from a combined analysis of 14 constitutive promoters and nine RBSs. The predicted growth rate from engineered *E. coli* with 9 different RBSs nearly equals to the actual growth rate (Fig. 1*d*) while prediction for engineered *E. coli* with 14 different constitutive promoters shows larger variance, though statistically not significant (Fig. 1*e*).

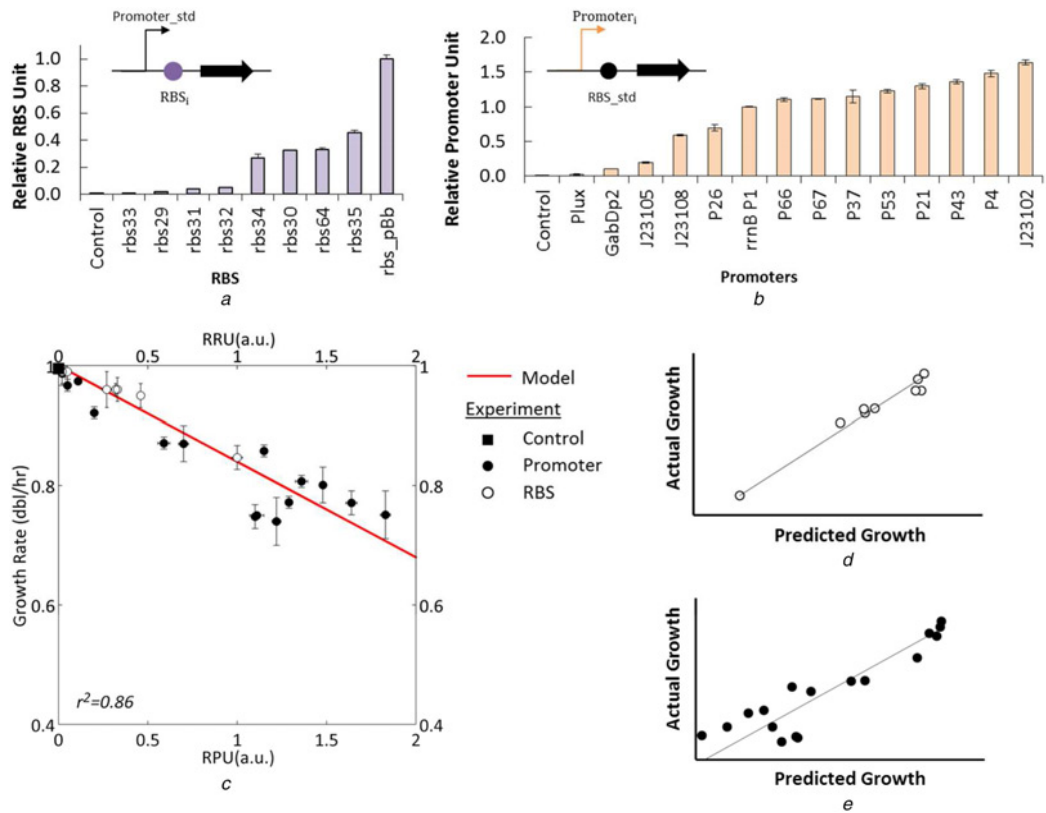

**Fig. 1** *Validation of growth reduction caused by single heterologous expression*
*a, b* Characterisation of standard biological parts including constitutive promoters and RBSs
*c* Heterologous expressions driven by these parts are modelled (red line) and measured (markers). Control sample refers to the wild type *E. coli* without genetic circuits expressing heterologous proteins
*d, e* Comparison between predicted and actual growth rate for each RBS and promoter

Our results show that the prediction is accurate under the two scenarios studied: (i) only varying promoter; and (ii) only varying RBS. Interestingly, varying RBS shows more accurate growth rate prediction (Figs. 1*d* and *e*), suggesting that the model works better with varied RBS than promoter in heterologous genetic circuits. We postulate that this could be because changing the promoter will affect both transcription and translation while changing RBS will mainly affect translation. Increasing promoter strength will lead to increase in mRNA production, and usage of polymerases and nucleotides. With an increase in mRNA, an increase in translation will likely to occur, resulting in increased amino acid usage. However, increasing RBS strengths will lead to increase in the translation rate of mRNA while the amount of mRNA will remain fairly constant, resulting in only increased amino acid usage but not nucleotide usage nor taxing on transcriptional machineries.

### 4.2 Significant correlation between predicted and observed protein abundance in a single expression cassette

It will be useful to predict the population-based heterologous protein abundance based upon growth rate because it corrects the dilution rate of heterologous proteins. We extend our initial model to predict overall heterologous protein abundance at any time point during the exponential growth phase, using biological parts characteristics (RPU and RRU) as input. Using RFP as reporter, the relationship between the RFP quantity and fluorescent signal was established (Fig. S5). This enables the conversion of RFP fluorescence signal (unit a.u) into protein abundance (molarity per OD600) by incorporating cell density. Using engineered *E. coli* carrying genetic circuit with reference parts (i.e. promoter *rrnB P1* with standard RBS and RFP gene), we experimentally estimated RFP abundance to be $1.7 \times 10^{-6}$ M per OD600 unit. Using this relation, we can calculate the population-based RFP abundance driven by different promoters of varying strengths when standard RBS was applied (Fig. 2*a*), and by different RBSs of varying strengths when a standard promoter was applied (Fig. 2*c*). Model prediction correlates well with the measured protein abundance (Figs. 2*b* and *d*) during exponential growth ($R^2 > 0.9$, *p*-value < 0.05), which is consistent with previous studies on modelling ribosome allocation between endogenous expression and transgenic expression [44, 45]. With a stronger promoter or RBS, this prediction clearly profiles the coherent and synchronous increase of heterologous protein with longer time due to burden. The results demonstrate that for a desired heterologous protein accumulation at a certain time or during a period (within exponential phase), our model could be used to select a suitable part: promoter or RBS, to achieve the desired protein abundance.

### 4.3 Model predicts protein abundance from two transgenic expression cassettes

Our model has shown to be able to predict host cellular growth and heterologous protein abundance when engineered *E. coli* expresses one heterologous protein by a promoter. Next, we investigate the validity of our model when engineered *E. coli* expresses two or more heterologous proteins. As input to our model, we presume that RPU is additive and together will lead to growth reduction of the host (Fig. 3*a*). To validate this hypothesis, we constructed and characterised a LasR-plasI QS circuit as a case study (Fig. 3*b*). The QS circuit has a typical gene circuit topology in which a constitutive promoter drives the production of a transcription factor (TF) (LasR protein) while an inducible promoter (plasI), which can be activated by the TF or its complex, expresses a reporter protein (RFP). As proposed in the model, the total RPU of the circuit ($RPU_{TOT}$) is represented as a sum of RPU of the constitutive promoter ($RPU_{constitutive}$) and the inducible promoter plasI, $RPU_{Inducible}$. To determine $RPU_{inducible}$, we characterise plasI

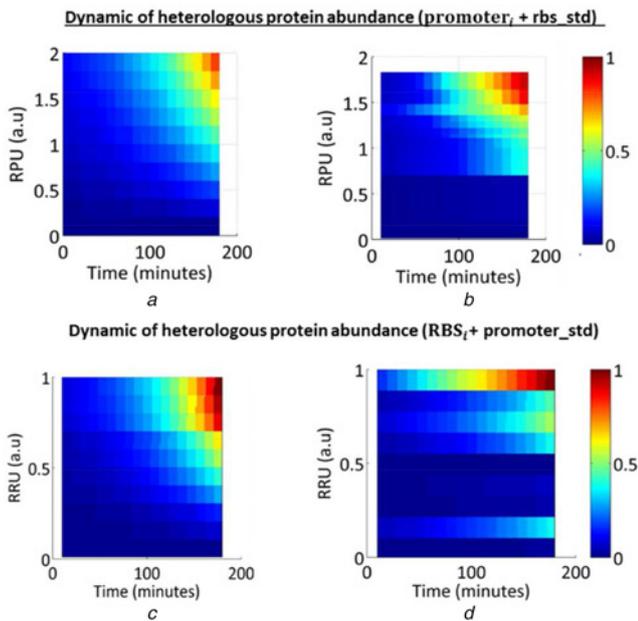

**Fig. 2** *Validation of heterologous protein abundance*

*a* Model proposed in this study predicts the population based protein abundance driven by various promoters

*b* Experimental result shows that these population based protein produced by genetic circuits with various promoters agrees with predicted protein abundance

*c* Model proposed in this study predicts the population based protein abundance driven by various RBSs

*d* Experimental result shows that these population based protein produced by genetic circuits with various RBSs agrees with predicted protein abundance with LasR protein being expressed under a standard constitutive promoter (*rrnB P1*). $RPU_{inducible}$ is represented as RFP abundance over a range of inducer (AHL) concentrations (Fig. 3*c*). With a fixed $RPU_{constitutive}$ and a gradual increase in $RPU_{inducible}$, we expect an increase in $RPU_{TOT}$ which will result in growth reduction. Experimental results show a growth rate reduction of 60% when $RPU_{TOT}$ increases to 3.70 (Fig. 3*d*). The experimental results strongly correlate with the model ($R^2 = 0.94$, *p*-value < 0.01), suggesting that the RPU is additive under the scenario studied.

There are other commonly used inducible promoters in synthetic biology, for example, pLlacO-1, pTet and pBAD, whose repressive expression would be relieved with the presence of inducers (Fig. 4*a* and Fig. S6). Similar to a plasI promoter, after characterising pLlacO-1, pTet and pBAD by determining $RPU_{inducible}$ over a range of inducers (Fig. 4*b*), respectively, we estimated $RPU_{constitutive}$ and applied our model to predict the growth rate and protein abundance for transgenic expression cassettes. Interestingly, we observed negligible growth reduction when inducers were not introduced (Figs. S6 and S7). Since zero induction demonstrates that inducible promoters (pLlacO-1, pTet and pBAD) are inactive, it mimics the situation that only repressors (Laclq, TetR and AraC) are expressed. Hence, we deduce that the cellular resource allocated to repressors' expression is negligible; in other words, $RPU_{constitutive}$ in the cassettes are very small (as shown in Fig. 4*a*). Therefore, we use $RPU_{inducible}$ to predict growth rates, which shows good correlation with observed growth rate (Fig. 4*c*, $R^2 = 0.91$, *p*-value < 0.01). Surprisingly, inducible promoters pTet and pBAD are much stronger than the previously studied 14 constitutive promoters (Fig. 1*b*), reaching a maximum $RPU_{inducible}$ of about 3.0. Even with stronger heterologous promoter, this model can predict the corresponding growth reduction. This suggests that the model is valid even when a strong heterologous transcription occurs. Furthermore, the population-based RFP abundance calculated by the model agrees well with measured RFP abundance (Fig. 4*d*, $R^2 > 0.90$, *p*-value < 0.01).

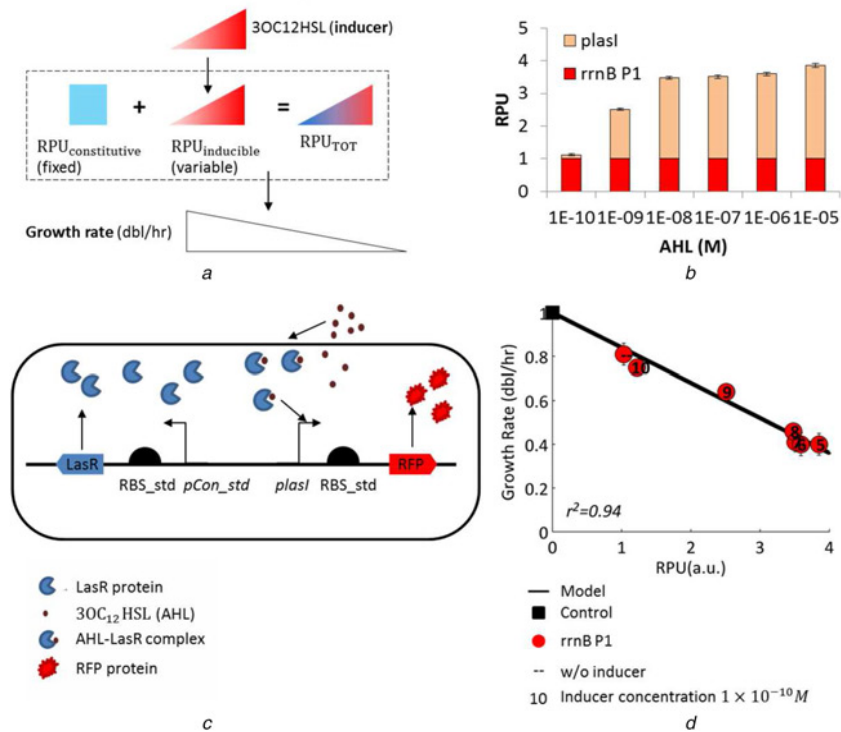

**Fig. 3** *Validation of growth in QS circuit*

*a* Two transgenic expression cassettes. Growth reduction is modelled by additive RPU (Equation S15)

*b* Schematic representation of QS circuits in engineered *E. coli*

*c* plasI promoter is characterised by AHL (inducer) with six different concentrations. RPU of this promoter reaches a plateau when AHL > $10^{-8}$ M. RPU of pCon_std (*rrnB P1*, Fig. 1*b*) is used to calculate $RPU_{TOT}$

*d* Validation of growth reduction. Numeric labels '5' to '10' refer to AHL concentrations from $10^{-5}$ to $10^{-10}$ M. Control sample refers to wild type *E. coli* without QS circuits expression LasR protein and RFP

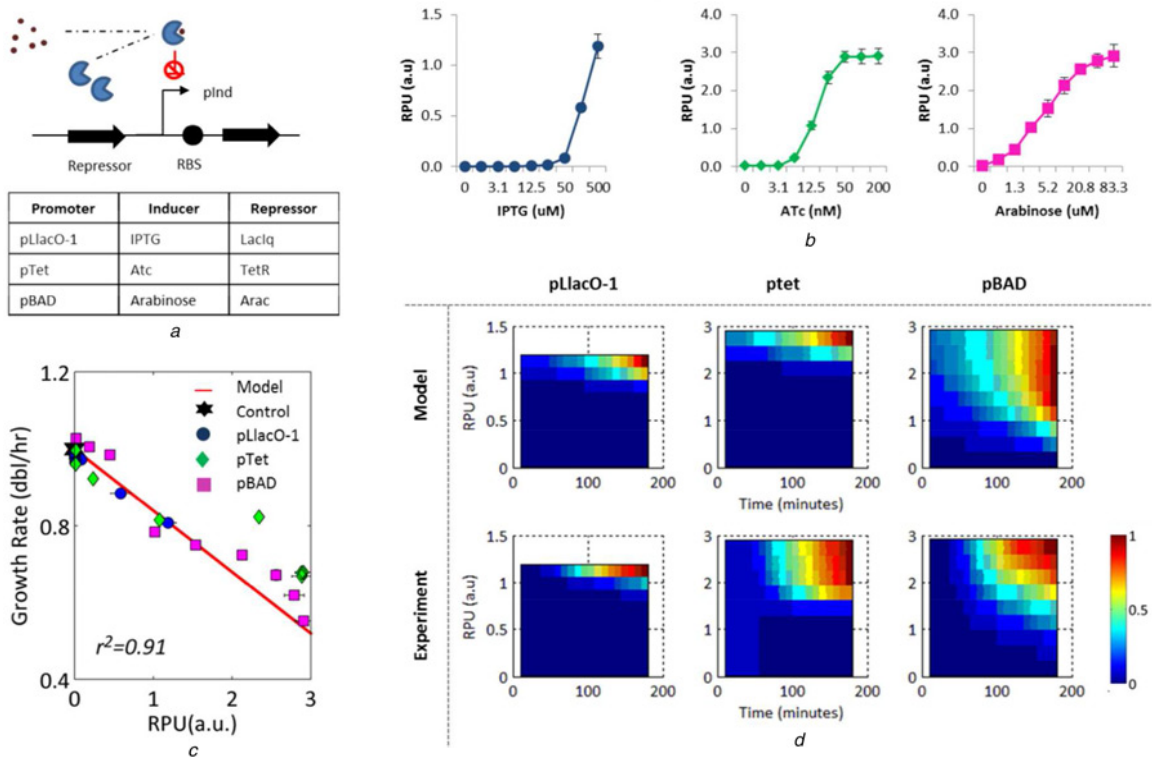

**Fig. 4** *Validation of growth reduction and protein abundance caused by inducible expression*

*a* Schematic representation of genetic circuits with three inducible promoters and repressors
*b* Characterisation of inducible promoters with varied concentrations of inducers
*c* Validation of growth reduction caused by three inducible expressions shown in (b). Control sample refers to the wild type *E. coli* without genetic circuits expressing heterologous proteins
*d* Comparison between predicted and measured abundance of heterologous proteins

The above validation shows that our model explains the correlation between host growth rate and the dynamic profile of heterologous protein abundance, which is consistent with optimal resource partitioning between endogenous expression and transgenic expression [27]. We find that considering growth rate clearly improves the prediction of total heterologous protein abundance (Fig. 5*a*) because the dynamic of population size is determined by the growth rate (Fig. S2). Interestingly, the model

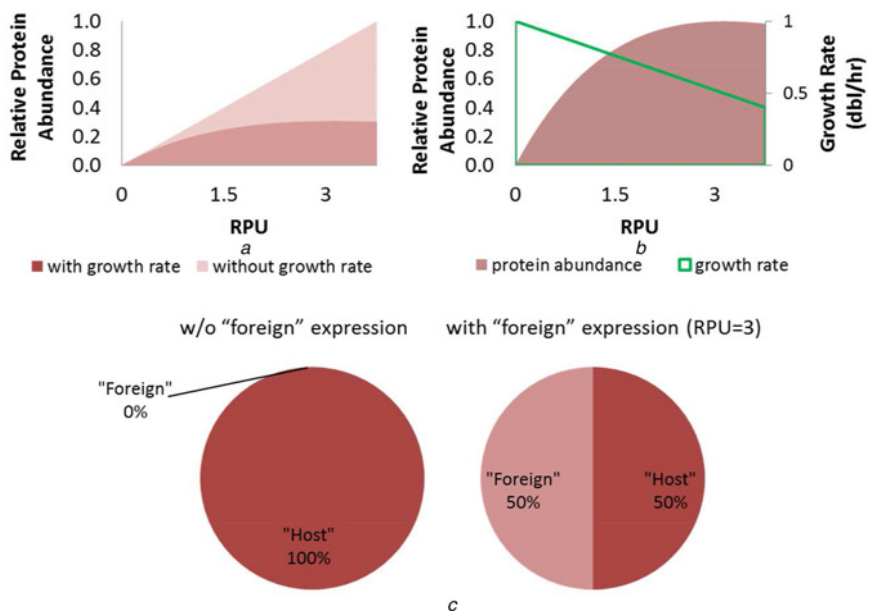

**Fig. 5** *Cellular resource sharing model facilitates the understanding of heterologous protein abundance*

*a* Model proposed in this study shows that total heterologous protein abundance increases less rapidly by considering growth rate which affects the heterologous protein dilution rate
*b* Model shows a linear reduction in growth rate with the increase in RPU. Both RPU and growth rate determine the total heterologous protein abundance which reaches a plateau when RPU > 3
*c* Example of resource allocation for wild type *E. coli* and engineered *E. coli* expressing heterologous protein with a RPU = 3

suggests that total heterologous protein reaches a plateau when RPU > 3 (Fig. 5a). This plateau may be due to the balance between increased RPU and reduced growth rate (Fig. 5b). When RPU equals to 3, our model suggests a 50%:50% cellular resources allocation between the host and 'foreign' genetic circuit (Fig. 5c). We modelled that there is a balance between glucose uptake and consumption rate. Since a constant glucose uptake rate in native and engineered *E. coli* is applied (Fig. S4), we postulate that glucose consumption rate remains unchanged in engineered *E. coli*. Together with the cellular resource allocation proposed in Fig. 5c, our model suggests a 50% reduction in cellular resource used by host, which agrees with ∼50% increase in cell dividing time ($\mu = 0.52$ dbl/h).

### 4.4 Our model is a potential tool for a priori optimisation of genetic circuits design

To demonstrate that our model aids in the optimisation of gene circuits, we revisited the LasR-*plasI* QS circuit and performed a proof-of-concept case study to find an optimal design. The objective is to derive a design that produces the maximum reporter protein (RFP) abundance using the library of characterised promoters available.

Our model suggests that RFP abundance is determined by both $RPU_{inducible}$ and host's growth rate which is affected by $RPU_{constitutive}$ and $RPU_{inducible}$. As plasI is activated by AHL-LasR complex, the RPU of plasI promoter depends on the amount of LasR and AHL present. The amount of LasR relies on the RPU of the constitutive promoter driving its expression. Fig. 4c shows that the fully activated plasI has a RPU of 2.50 when AHL concentration is more than $10^{-8}$ M. We noted LasR abundance in this genetic circuit ($1.7 \times 10^{-8}$ M as estimated using the RPU of *rrnB P1* promoter, which drives LasR expression) is higher than AHL concentration when plasI is fully activated. Taken together, the QS circuits (in Fig. 4b) could have overproduced LasR protein, resulting in a suboptimal use of cellular resource. Hence, a novel design which could produce LasR at a more optimal level might achieve an increase in RFP abundance via a less reduced growth rate while maintaining plasI activity.

To determine such a design, we derived a trend of overall RFP abundance when $RPU_{constitutive}$ was varied over a range from 0 to 2 (Fig. 6a). In the derivation, because plasI is activated by AHL-LasR complex, we postulated that AHL-LasR complex recruits 1:1 AHL and LasR protein, respectively. The trend was obtained using the model and a relationship between $RPU_{inducible}$ and $RPU_{constitutive}$ established (Fig. 6b). Interestingly, Fig. 6a

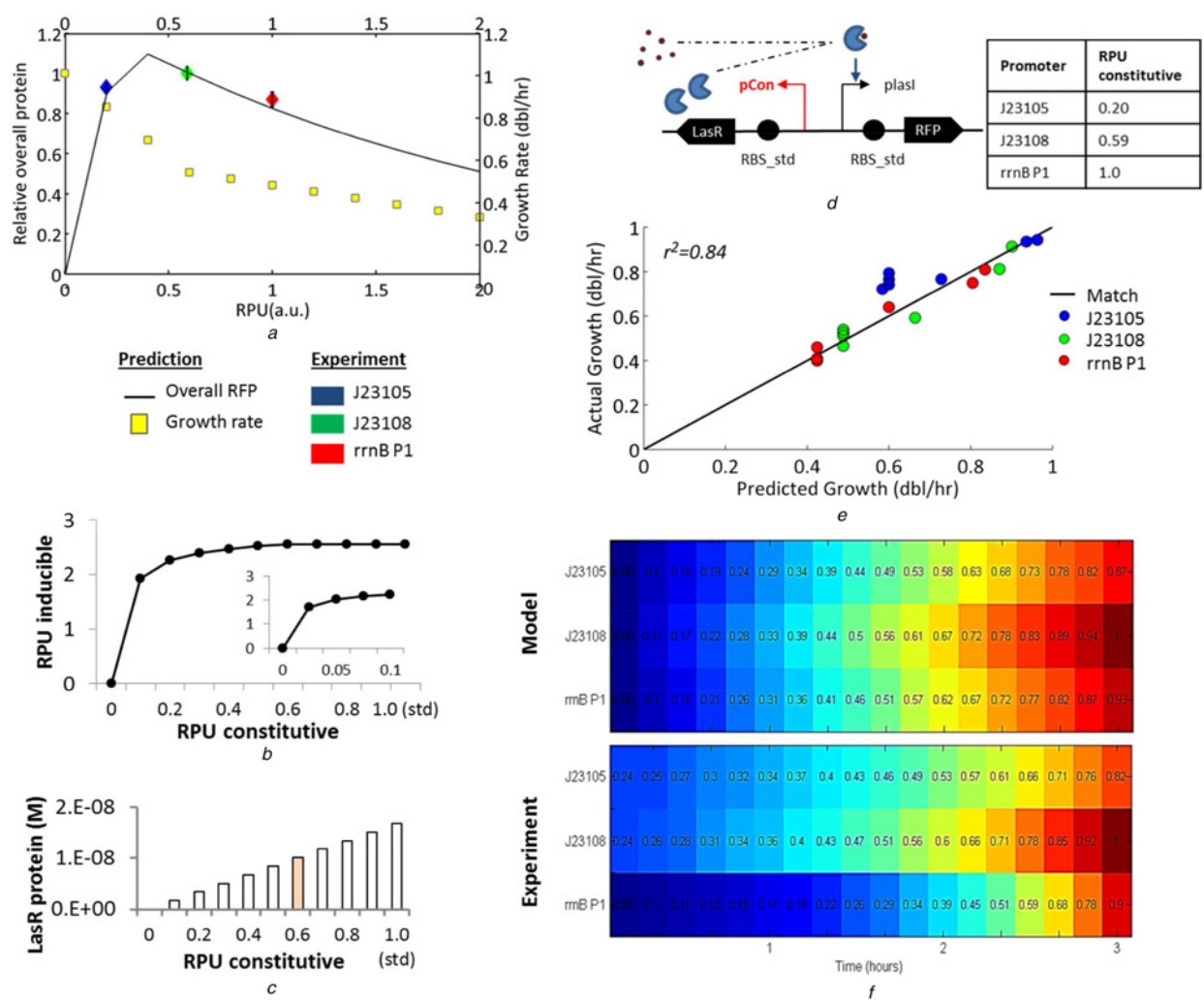

**Fig. 6** *Model aids in the optimisation of genetic circuits*
*a* For QS expression (AHL-plasI), when plasI is fully activated (Fig. 5c), our model shows that the growth reduction and heterologous protein abundance are affected by the LasR production (driven by various constitutive promoters). Experimental measured protein abundance agrees with model prediction at 3 h
*b* Model derived relationship between the strength of constitutive promoter (driving LasR production) and max ability of inducible promoter (plasI, activated by LasR-AHL complex)
*c* Model predicted LasR protein abundance when constitutive promoter with RPU ranging from 0 to 1 is used
*d* Schematic representation of candidate QS constructs with three different constitutive promoters
*e* Validation of growth reduction in engineered *E. coli* carrying each candidate QS construct in (*d*). For each construct, AHL concentrations ranging from $10^{-10}$ to $10^{-5}$ M are introduced
*f* For each candidate QS construct, maximum plasI ability is reached when AHL > $10^{-8}$ M (Supporting information S8). Model predicted total RFP abundance agrees with the experimental measurement of RFP during exponential growth phase suggests that there is an optimal $RPU_{constitutive}$ (∼0.4) that can be used to achieve the maximum RFP abundance. RFP abundance first increases when $RPU_{constitutive}$ increases from 0 to 0.4. When $RPU_{constitutive}$ is further increased beyond 0.4, RFP abundance starts to decrease, resulting from a more rapid decrease in growth rate. This is due to the larger burden with a larger production of lasR when $RPU_{constitutive}$ increases.

We estimated LasR abundance over a range of RPU (Fig. 6c). The results show that LasR abundance would reach $10^{-8}$ M at a RPU of 0.6, suggesting that using promoters with RPU more than 0.6 would overproduce LasR and saturate plasI when AHL is more than $10^{-8}$ M. Based on the results, we performed combinatorial studies in which QS circuits with varied promoters driving LasR transcription were constructed (Fig. 6d). As our library of promoters did not have a promoter with RPU of 0.4, we chose constitutive promoters BBa_J23105 and BBa_J23108 with RPU of 0.20 and 0.59, respectively. These two promoters have RPU closest to 0.4. Thus, together with reference promoter $rrnB \ P1$ (RPU = 1.0), we constructed three QS constructs. Each of the constructs only differs in the constitutive promoter driving the expression of LasR. We expected a higher RFP abundance to be produced by genetic circuit with BBa_23108 as its RPU is closer to optimal $RPU_{constitutive}$ than the other two promoters.

The measured growth rates of the three engineered E. coli (each carrying one of three QS constructs, respectively) over a range of AHL concentrations show good correlation between the model predicted growth rates (Fig. 6e, $R^2 > 0.85$, $p$-value < 0.01). This further demonstrates that the additive property of RPU proposed in the model holds in this. As expected, among the three genetic circuits, the most abundant RFP was produced by BBa_J23108 driven LasR (Fig. 6a). The measured and predicted RFP abundance over time during exponential phase also show good correlation (Fig. 6f, $R^2 > 0.90$). These findings imply that balancing TF abundance improves reporter protein production in our QS circuits. This suggests that our model could be used to find the optimal TF abundance to aid in genetic circuits design. In short, insufficient amount of TF limits the inducible promoter activity while overproduced TF might stress the host too much and cause a further reduced growth rate which will then affect the abundance of reporter protein.

In this study, we present two models to describe growth rate reduction due to cellular resource sharing between host circuitry and 'foreign' genetic circuits expressing heterologous gene. Reduction in growth rate has subtle effect on the performance of genetic circuits as it varies protein dilution rate. Characteristics of biological parts, such as RPU and RRU, are used as inputs to model growth reduction, to enable the model to be used during the design of genetic circuits. Our results corroborate previous studies on inter-dependence of host cell growth and genetic circuit's behaviour [11] (Fig. S7), showing the importance in considering growth impact during the tuning of genetic circuits to achieve maximum protein abundance. In addition, our results suggest that growth reduction predictions correlate well (average $R^2$∼0.9) with experimental data, and is comparable to the correlations in other studies using more parameters [11, 16, 17, 45]. This suggests that RPU and RRU can be used as high-level, coarse-gained abstractions to the partitioning of cellular resources towards heterologous expression under the assumption of optimal resource allocation [24, 25]. This also suggests that our models can be handy tools to estimate growth reduction from the impact of cellular resource usage of heterologous expressions as the models require part characteristics, which are readily available from standard parts characterisation experiments [29].

Our promoter model was experimentally validated up to RPU of 3.7, accompanying with a growth reduction of up to 60%. Further work may include the upper bound identification of RPU. This maximum RPU may assist in understanding the constraint of host E. coli in supporting heterologous expression.

Our models study the allocation of cellular resource during exponential growth phase. Cellular component synthesis and synthetic promoter activity in other phases, like lag and stationary phases, are known to be different. Hence, application of our model is limited to exponential growth phase.

Our models assume that the products from the genetic circuits are non-toxic to the host. The reduction in growth is mainly due to the depletion of cellular resource available for the host to grow. It is important to note that the model assumes a particularly experimental setting (methods). As parts characterisation varied with experimental setting (strains, vector backbones, medium), the proposed model needs to be recalibrated for different experimental setting.

Our models consider promoter and RBS independently, despite experimental evaluation with two promoters and demonstrating the additive effect of promoter strengths on growth reduction. Future work can consolidate these two models into one by considering the possible interaction between promoter and RBS using combinatorial circuits, and in single and multiple expression cassettes.

## 5 Acknowledgment

The authors appreciate the support from Associate Professor Matthew Wook Chang. They thank Dr Wendy Chen and Dr Tat-Ming Lo for constructing p-series constitutive promoters.

## 6 Data statement

Experimentally observed and predicted growth rates originated in this study are listed in Tables S3, S4 and Extension S1 of the Supplementary Information.

## 7 Author contributions

HW and CLP conceived the project. HW developed the model and analysed the data, with significant contribution from ML. TZC performed the experiments. HW, CLP and ML prepared the manuscript. All authors have given approval to the final version of the manuscript.

## 8 Funding sources

Work on this project is supported by Singapore MOE Academic Research Fund Tier 2 Grant (AcRF ARC43/13).

## 9 Declaration of interests

The authors declare no competing financial interests.

# Supporting Information

# Contents

## Figure S1: Cell geometry calculation

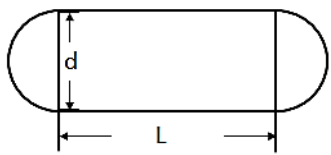

Figure S1: *E. coli* cell geometry

- Surface area: $A^H = \pi(\frac{d^2}{2} + Ld)$

- Volume: $V^H = \pi(\frac{d^3}{6} + L\frac{d^2}{4})$

- Volume-related area:

$$A^H = \frac{4V^H}{d} - \frac{\pi d^2}{6}$$

$$\Big\downarrow V^H = \frac{M^H}{\rho}$$

$$A^H = \frac{4M^H}{\rho d} - \frac{\pi d^2}{6} \approx \frac{4M^H}{\rho d}$$

# Figure S2: Heterologous protein abundance

Reference to [1].

- RPU is accessible by comparing the expression level in fluorescence intensity (reporter protein is fluorescence protein)

$$RPU = \frac{\text{fluorescence intensity of target protein}}{\text{fluorescence intensity of reference protein}}$$

- RPU is measured at exponential growth

$$RPU = \alpha_P^V / \alpha_{ref}^V = \mu \times P_P^V / \mu_{ref} \times P_{ref}^V$$

- Heterologous protein abundance

$$P_P^V = \frac{RPU \times \mu_{ref} \times P_{ref}}{\mu}$$

# Figure S3: Overall heterologous protein

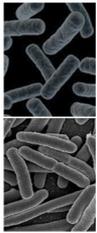

$t = t_0$
- $N = N_0$
- $P_{all} = P^V \times N_0$

$t = t_1$
- $dN/dt = \mu N$
- $dP_{all}/dt = P^V \times dN/dt$

Figure S3: *E. coli* cell population growth (adapted from SEM images)

- Heterologous protein per bacterial cell maintains

$$P^V = P_{ref} \times RPU(\text{or, RRU})$$

- Population growth

$$N = N_0 \times 2^{t/td}$$

- Doubling time is derived from growth rate

$$td = \ln 2 / \mu$$

# Figure S4: Glucose uptake rate study

**Calculation**

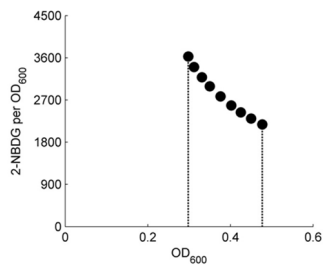

Figure S4A: Glucose analog (2-NBDG) residues per OD

$$AUC = \int_{OD_1}^{OD_2} \frac{NBDG_{residues}}{OD} dOD$$

$$AUC = \int_{t_1}^{t_2} \int_{OD_1}^{OD_2} \frac{NBDG_{t=0} - C_{uptake} \times OD \times dt}{OD} dOD$$

$$V_{uptake}^H = \frac{NBDG_{t=0} \times (\ln(OD_2) - \ln(OD_1)) - AUC}{(t_2 - t_1) \times (OD_2 - OD_1)}$$

$NBDG_{residues}$ : 2-NBDG residues at time $t$ with cell number $N$
$N$ : $0.3 \times OD(t) \times 10^8$, 0.3 refers to sample volume 300ul
$V_{uptake}^H$ : glucose uptake rate, equals to 2-NBDG uptake rate
$AUC$ : area under the curve

**Result**

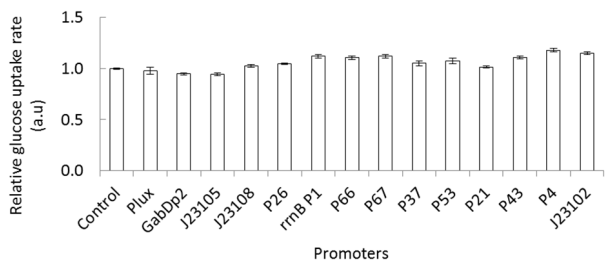

Figure S4B: Constant glucose uptake rate in wildtype (control) and engineered *E. coli* carrying genetic circuits expressing RFP driven by 14 types of constitutive promoters

# Figure S5: RFP quantification

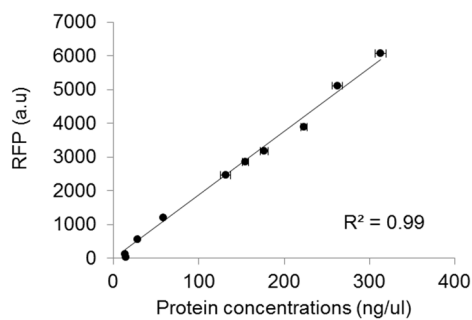

Figure S5: protein quantification curve

- RFP produced by vivo-reference genetic circuit during exponential phase is **970 (a.u),** corresponds to **51 ng/ul**

- RFP molecular weight is 30kDa, convert **51 ng/ul** into **1.7uM**

⇨ $P_{ref} = 1.7 \times 10^{-6}\ M/OD$

# Figure S6: Induction system

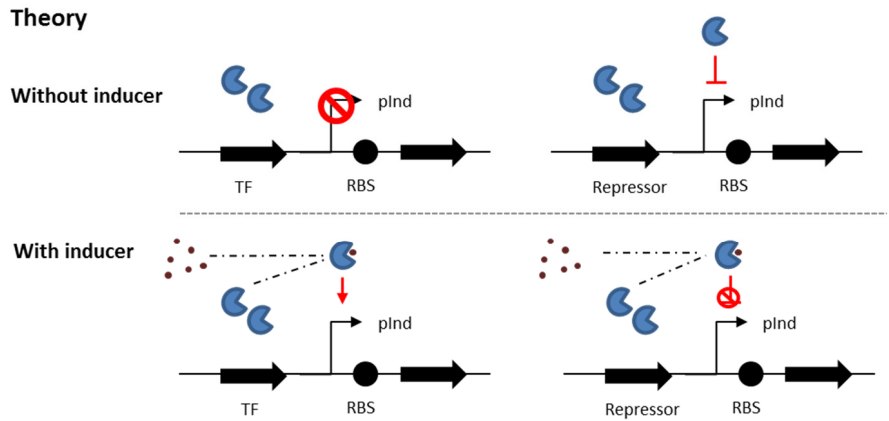

Figure S6: Inducible expression regulated by transcriptional factor (TF) or repressor.

# Figure S7: Negligible burden caused by repressor production

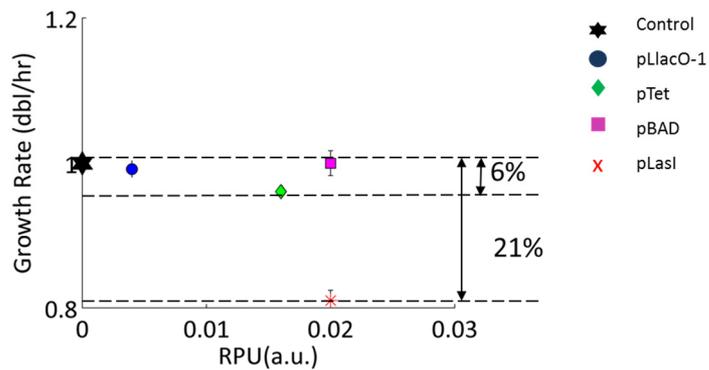

Figure S7: Negligible burden caused by repressor production. In the genetic circuits studied in Figure 3A and Figure 5A, only repressor would be produced with the absence of inducer (Figure S6), and further results in growth reduction. Comparing with the significant growth reduction (21%) caused by LasR protein production (driven by rrnB P1 promoter), we observe negligible growth reduction (<6%) caused by the production of LacIq, TetR and Arac proteins (aiding in the activation of pLlacO-1, pTet and pBAD promoter respectively, Figure 3A).

# Figure S8: Model comparison

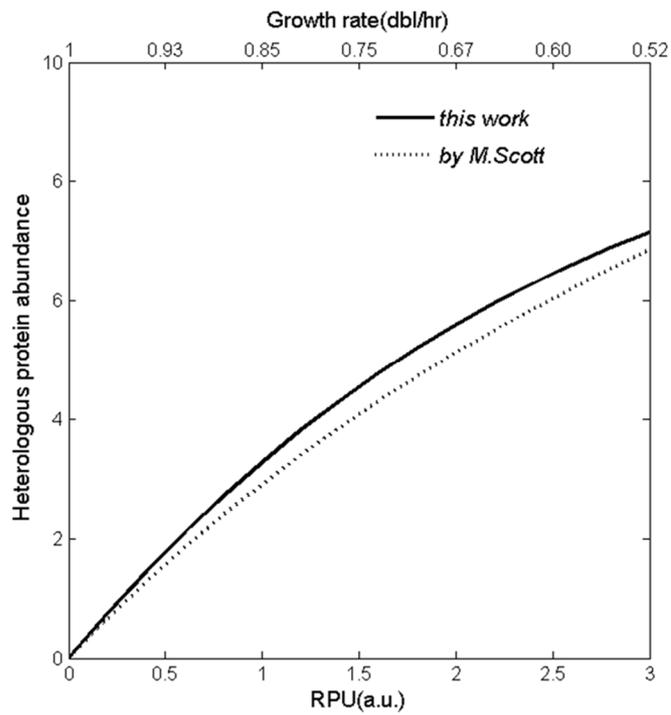

Figure S8: Comparison of predicted heterologous abundance between our model and Scott *et al.* [2]

# Figure S9: Predicted RPUTOT of three quorum sensing circuit

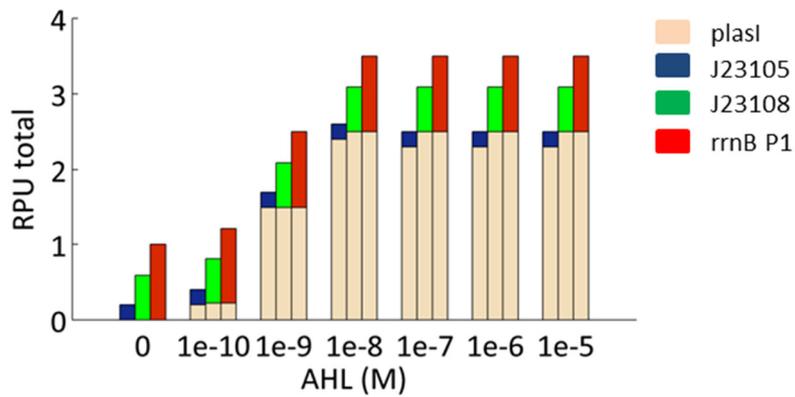

Figure S9: predicted RPUTOT of three quorum-sensing circuits used for the validation of growth reduction (Figure 6E) and total RFP production (Figure 6F)

## Description S1: Assembly of model

Given superscript H (such as $V_{consumption}^{H}$) to represent host variables and superscript P (such as $V_{consumption}^{P}$) to represent plasmid / genetic circuit variables,

$$V_{updake}^{H} = V_{consumption}^{H} + V_{consumption}^{P}$$

Can be resolved as

$$2 \times 10^{21} \times M^{H} = (M_{protein}^{H} + E_{protein}^{H}) + \\ (M_{RNA}^{H} + E_{RNA}^{H}) + \\ (M_{lipid}^{H} + E_{lipid}^{H}) + \\ (M_{protein}^{P} + E_{protein}^{P})$$

$$2 \times 10^{21} \times M^{H} = (\varphi_{protein}^{H} + 1) E_{protein}^{H} + \\ (\varphi_{RNA}^{H} + 1) E_{RNA}^{H} + \\ (\varphi_{lipid}^{H} + 1) E_{lipid}^{H} + \\ (\varphi_{protein}^{P} + 1) E_{protein}^{P}$$

$$2 \times 10^{21} \times M^{H} = (\varphi_{protein}^{H} + 1)(\varepsilon \times c_{protein}^{H} \times \alpha_{protein}^{H}) + \\ (\varphi_{RNA}^{H} + 1)(\varepsilon \times c_{RNA}^{H} \times \alpha_{RNA}^{H}) + \\ (\varphi_{lipid}^{H} + 1)(\varepsilon \times c_{lipid}^{H} \times \alpha_{lipid}^{H}) + \\ (\varphi_{protein}^{P} + 1)(\varepsilon \times c_{protein}^{P} \times \alpha_{protein}^{P})$$

$$2 \times 10^{21} \times M^{H} = (\varphi_{protein}^{H} + 1)(\varepsilon \times c_{protein}^{H} \times \mu \times P_{protein}^{H}) + \\ (\varphi_{RNA}^{H} + 1)(\varepsilon \times c_{RNA}^{H} \times \mu \times P_{RNA}^{H}) + \\ (\varphi_{lipid}^{H} + 1)(\varepsilon \times c_{lipid}^{H} \times \mu \times P_{lipid}^{H}) + \\ (\varphi_{protein}^{P} + 1)(\varepsilon \times c_{protein}^{P} \times \mu \times P_{protein}^{P})$$

$$2 \times 10^{21} \times M^{H} = (\varphi_{protein}^{H} + 1)\left(\varepsilon \times c_{protein}^{H} \times \mu \times \frac{\sigma_{protein}^{H} \times M^{H}}{MW_{protein}^{H}} \times NA\right) + \\ (\varphi_{RNA}^{H} + 1)\left(\varepsilon \times c_{RNA}^{H} \times \mu \times \frac{\sigma_{RNA}^{H} \times M^{H}}{MW_{RNA}^{H}} \times NA\right) + \\ (\varphi_{lipid}^{H} + 1)\left(\varepsilon \times c_{lipid}^{H} \times \mu \times \frac{4 \times 0.5 \times A^{H}}{A_{lipid}^{H}}\right) + \\ (\varphi_{protein}^{P} + 1)(\varepsilon \times c_{protein}^{P} \times RPU \times \mu_{std} \times P_{std})$$

$$2 \times 10^{21} \times M^{H} = \mu \left[(\varphi_{protein}^{H} + 1)\left(\varepsilon \times c_{protein}^{H} \times \frac{\sigma_{protein}^{H} \times M^{H}}{MW_{protein}^{H}} \times NA\right) + \\ (\varphi_{RNA}^{H} + 1)\left(\varepsilon \times c_{RNA}^{H} \times \frac{\sigma_{RNA}^{H} \times M^{H}}{MW_{RNA}^{H}} \times NA\right) + \\ (\varphi_{lipid}^{H} + 1)\left(\varepsilon \times c_{lipid}^{H} \times \frac{4 \times 0.5 \times A^{H}}{A_{lipid}^{H}}\right)\right] + \\ (\varphi_{protein}^{P} + 1)(\varepsilon \times c_{protein}^{P} \times RPU \times \mu_{std} \times P_{std})$$

$$(2 \times 10^{21} \times M^H) -$$
$$\mu \left[ (\varphi_{protein}^H + 1) \left( \varepsilon \times c_{protein}^H \times \frac{\sigma_{protein}^H \times M^H}{MW_{protein}^H} \times NA \right) + (\varphi_{RNA}^H + 1) \left( \varepsilon \times c_{RNA}^H \times \frac{\sigma_{RNA}^H \times M^H}{MW_{RNA}^H} \times NA \right) \right.$$
$$\left. + (\varphi_{lipid}^H + 1) \left( \varepsilon \times c_{lipid}^H \times \frac{4 \times 0.5 \times A^H}{A_{lipid}^H} \right) \right]$$
$$= (\varphi_{protein}^P + 1)(\varepsilon \times c_{protein}^P \times RPU \times \mu_{std} \times P_{std})$$

# Table S1: Parameters used in biophysical model

| Cellular component | Protein | RNA | DNA | Lipid |
|---|---|---|---|---|
| Ratio between building material and energy cost ($\varphi$) [3] | 4 | 9.5 | 9.5 | 7 |
| Energy cost constant per unit length[a] | 5.2 | 70 | 70 | 180 |
| Molecular unit length[b] | 300 | 380 | | 1 |
| % of total dry weight ($\delta$) [2] | 55 | 16.5 | 3.1 | 12.5 |
| Half-life | >hours | Stable[c] | Stable | Stable[C] |
| Molecular weight (MW, unit kDa) | 30 | 440 | - | - |
| Molecular area ($A_{lip}$) [3] | - | - | - | 5x10$^{-13}$ |

[a,b]: used to calculate energy cost consant ($c_i$) = energy cost constant per unit length × molecular unit length [4]

[c]: stable means the degradation rate is negligible [5-7]

# Table S2: Constitutive promoters in this study

| Name | Catalogue | Sequence |
|---|---|---|
| BBa_K091157 (Plux) | *E. coli* pLux/Las hybrid promoter | CTATCTCATTTGCTAGTATAGTCGAATAAA |
| gabDP2 | *E. coli* gabDT operon | GAGATTTTGGGCTCGTCGGGGATTCGCCGGG TGCTGCAAAACCATCTACGCTCAGGACTGGG CGAGATGAAAAACTCGCTG |
| BBa_J23105 | *E. coli* $\sigma_{70}$ promoter | GGCTAGCTCAGTCCTAGGTACTATGCTAGC |
| BBa_J23108 | *E. coli* $\sigma_{70}$ promoter | AGCTAGCTCAGTCCTAGGTATAATGCTAGC |
| p26 | *E. coli* $\sigma_{70}$ promoter | ATTAAAAGAATAAAATTCTTGACATATTAAG CACGATATGATATAATATCTGAGGA |
| BBa_K112118 | *E. coli* rrnB P1 promoter | ATAAATGCTTGACTCTGTAGCGGGAAGGCG |
| p66 | *E. coli* $\sigma_{70}$ promoter | GTATATATTAAAACATTCTTGACATCTTGAA ACAAATATGATATAATAGCAATATAT |
| p67 | *E. coli* $\sigma_{70}$ promoter | ATAAAATTAAAAAATTCTTGACATCTTCAC AAAAATATGATATAATGCACCTATAA |
| p37 | *E. coli* $\sigma_{70}$ promoter | GTATAAAGAAAAACTTCTTGACATATTAGA AAAAATATGATATAATGAAAAAATAA |
| p53 | *E. coli* $\sigma_{70}$ promoter | GTTTAATGAAAATCGTTCTTGACATGTTAAA TCAAATATGATATAATAAAAAAATAA |
| p21 | *E. coli* $\sigma_{70}$ promoter | CTAGTTTAGATTAAATTCTTGACATATTAAA AAAATTATGATATAATTCAAATATAT |
| p43 | *E. coli* $\sigma_{70}$ promoter | ATAAAGAACACAAAGTTCTTGACATATTAAA AACAATATGATATAATAAAAAAATAA |
| p4 | *E. coli* $\sigma_{70}$ promoter | ATAGACAAGAAAAATTCTTGACATATTAA AAACCATATGATATAATAAAAGCATAA |
| BBa_J23102 | *E. coli* $\sigma_{70}$ promoter | TTGACAGCTAGCTCAGTCCTAGGTACTGTGC TAGC |

# Table S3: Predicted vs observed growth rate (1 cassette)

| Experiment | RPU | RRU | Observed Growth (/hr) | Predicted Growth (/hr) | % Error |
|---|---|---|---|---|---|
| Vary promoter | 0.02 | 1 | 0.987 | 0.997 | 0.99 |
| Vary promoter | 0.11 | 1 | 0.974 | 0.982 | 0.86 |
| Vary promoter | 0.20 | 1 | 0.921 | 0.968 | 5.10 |
| Vary promoter | 0.59 | 1 | 0.871 | 0.906 | 3.97 |
| Vary promoter | 0.70 | 1 | 0.870 | 0.888 | 2.07 |
| Vary promoter | 1.00 | 1 | 0.847 | 0.840 | 0.83 |
| Vary promoter | 1.10 | 1 | 0.748 | 0.824 | 10.16 |
| Vary promoter | 1.11 | 1 | 0.750 | 0.822 | 9.65 |
| Vary promoter | 1.15 | 1 | 0.858 | 0.816 | 4.90 |
| Vary promoter | 1.22 | 1 | 0.740 | 0.805 | 8.76 |
| Vary promoter | 1.36 | 1 | 0.807 | 0.782 | 3.05 |
| Vary promoter | 1.48 | 1 | 0.801 | 0.763 | 4.72 |
| Vary promoter | 1.64 | 1 | 0.771 | 0.738 | 4.33 |
| Vary promoter | 1.83 | 1 | 0.751 | 0.707 | 5.83 |
| Vary RBS | 1 | 0.0004 | 1.010 | 1.000 | 1.00 |
| Vary RBS | 1 | 0.020 | 0.990 | 0.997 | 0.69 |
| Vary RBS | 1 | 0.040 | 1.000 | 0.994 | 0.64 |
| Vary RBS | 1 | 0.050 | 0.990 | 0.992 | 0.20 |
| Vary RBS | 1 | 0.270 | 0.960 | 0.957 | 0.33 |
| Vary RBS | 1 | 0.320 | 0.960 | 0.949 | 1.17 |
| Vary RBS | 1 | 0.330 | 0.960 | 0.947 | 1.33 |
| Vary RBS | 1 | 0.460 | 0.950 | 0.926 | 2.48 |
| Vary RBS | 1 | 1.000 | 0.847 | 0.840 | 0.83 |

These results demonstrated that growth rate inversely varies with promoter or RBS strengths.

For promoter variation (n = 14),
- Mean % error = 4.66%
- Standard error of % error = 0.832%

For RBS variation (n = 9),
- Mean % error = 0.96%
- Standard error of % error = 0.226%

# Table S4: Predicted vs observed growth rate (2 cassettes)

| RPU #1 | RPU #2 | Actual Growth (/hr) | Predicted Growth (/hr) | % Error |
|---|---|---|---|---|
| 0.59 | 2.10 | 0.599 | 0.570 | 4.91 |
| 0.59 | 0.81 | 0.812 | 0.776 | 4.43 |
| 0.20 | 0.52 | 0.935 | 0.885 | 5.34 |
| 0.20 | 1.71 | 0.766 | 0.694 | 9.38 |

Both cassettes are constructed using strong RBS (RRU of 1). The mean % error is 6.02% with a standard error of 1.137%. Although the mean % error between actual growth and predicted growth is higher compared to that of 1 cassette (6.02% error in 2 cassettes vs 4.66% error in 1 cassette; Table S3), this difference is not significant (p-value = 0.37).

# Extension S1: Preliminary work in combining promoter and RBS models

Preliminary work to combine promoter model and RBS model, developed in this study, is attempted.

When RBS of RRU of 1 was used (Table S3), promoters of 1.64 and 1.22 RPUs showed a growth rate of 0.771 and 0.740 divisions per hour were observed respectively. The predicted growth rates are 0.738 (4.33% error in prediction) and 0.805 (8.75% error in prediction) divisions per hour respectively. When promoter of 1 RPU was used with RRU of 0.32 (Table S3), a growth rate of 0.960 divisions per hour was observed and the predicted growth rate is 0.949 (1.17% error in prediction) divisions per hour.

Using a RBS of relative RBS strength of 0.32, two single cassette constructs using promoters of 1.64 and 1.22 RPUs were performed, and growth rate of 0.750 and 0.808 divisions per hour were observed respectively. These observed growth rates are closer to their respective promoter growth rates (0.750 vs 0.771, and 0.808 vs 0.740) when a much stronger RBS (RRU of 1) compared to the observed growth rate of 0.960 when RRU of 0.32 (with RPU of 1) was used. Less than 2% prediction errors are observed when our promoter model is used compared to more than 16% prediction errors are observed when our RBS model is used.

This suggests that our promoter model may provide a better estimate to growth reduction. Promoter strength (RPU) determines the amount of mRNA transcript and RBS strength (RRU) determines the amount of peptide chains from the amount of mRNA transcript. Hence, it is plausible to view RBS as a multiplier to the amount of mRNA transcript; thereby, the impact of RBSes may be reduced when compared to the impact of promoters.

This preliminary work to combine our promoter and RBS model is extremely limited with only 2 cases. Hence, significant of future work is needed for a thorough investigation.